\documentclass[letter]{jpconf}

\newcommand{\ls}{\ensuremath{l_s}} 

\def\p{\partial}

\def\expec#1{\langle #1 \rangle}

\newcommand{\cF}{{\mathcal{F}}}

\newcommand{\cL}{\mathcal{L}}

\newcommand{\cN}{{\mathcal{N}}}
\newcommand{\cO}{{\mathcal{O}}}

\newcommand{\tret}{{t_{\mbox{\scriptsize ret}}}}
\newcommand{\tx}{{\tilde{x}}}
\newcommand{\ttau}{{\tilde{\tau}}}

\begin{document}

\title{Radiation Damping in a Non-Abelian Strongly-Coupled Gauge Theory}


\author{Mariano Chernicoff$^{\dagger}$, J.~Antonio Garc\'{\i}a$^{\star}$ and Alberto G\"uijosa$^{\star}$}
  \address{$\dagger$ Departament de F\'\i sica Fonamental,
Universitat de Barcelona,
Marti i Franqu\`es 1, E-08028 Barcelona, Spain,
Email: {\tt mchernicoff@ub.edu}\\
${\star}$ Departamento de F\'{\i}sica de Altas Energ\'{\i}as, Instituto de Ciencias
  Nucleares \\
Universidad Nacional Aut\'onoma de M\'exico, Apdo. Postal 70-543,
M\'exico D.F. 04510, M\'exico\\
Email: {\tt garcia, alberto@nucleares.unam.mx}}

\begin{abstract}
 We study a `dressed' or `composite'
quark in strongly-coupled $\mathcal{N}=4$ super-Yang-Mills (SYM), making use of the AdS/CFT
correspondence. We show that the standard string dynamics nicely captures the physics of the quark
and its surrounding quantum non-Abelian field configuration, making it possible to derive a relativistic
equation of motion that incorporates the effects of radiation damping. {}From this equation one can
deduce a non-standard dispersion relation for the composite quark, as well as a Lorentz covariant
formula for its rate of radiation.
\end{abstract}


\section{Motivation}

A considerable effort has been invested in the study of strongly-coupled thermal
non-Abelian plasmas by means of the AdS/CFT correspondence \cite{malda,gkpw,magoo}.
The jet quenching observed at RHIC indicates that partons crossing the quark-gluon plasma experience significant energy loss. This question has been studied in a strongly-coupled $\mathcal{N}=4$ SYM plasma for a variety of probes, including quarks \cite{hkkky,gubser,quark}, mesons \cite{meson}, baryons \cite{draggluon,alr}, gluons \cite{draggluon,ggpr} and $k$-quarks \cite{draggluon}. Here we will analyze quark energy loss \emph{in vacuum}, which for a strongly-coupled non-Abelian gauge theory is interesting already in itself, and is in addition helpful for developing intuition that might extend to the finite temperature case \cite{dragtime,vacinplasma}.

 We expect an accelerating quark to radiate, and to experience a damping force due to the emitted radiation. In the context of
classical electrodynamics, and for a non-relativistic electron modeled as a
vanishingly small charge distribution, this effect is incorporated in the classic Abraham-Lorentz equation
\cite{abraham,lorentz} \begin{equation}\label{al}
m\left({d^2 \vec{x}\over dt^2}-t_e{d^3 \vec{x}\over dt^3}\right)=\vec{F}~.
 \end{equation}
The second term in the left-hand side is the damping term, which is seen to be associated with a characteristic timescale $t_e\equiv 2e^2/3mc^3$, set by the classical electron radius. The search for a Lorentz-covariant version
of (\ref{al}) led to the (Abraham-)Lorentz-Dirac equation \cite{dirac},
\begin{equation}\label{ald}
m\left({d^2 x^{\mu}\over d\tau^2}-t_e\left[
{d^3 x^{\mu}\over d\tau^3}-{1\over c^2}{d^2 x_{\nu}\over d\tau^2}\,
{d^2 x^{\nu}\over d\tau^2}{dx^{\mu}\over d\tau}\right]\right)=\cF^{\mu}~,
\end{equation}
with
$\tau$ the proper time and $\cF^{\mu}\equiv\gamma(\vec{F}\cdot\vec{v}/c,\vec{F})$
the 4-force.
The second term within the square brackets
is the negative of the rate at which 4-momentum is carried away from the charge by radiation (as given by the covariant Lienard formula), so it is only this term that can properly be called radiation reaction. The first term within the square brackets, usually called the Schott term, and whose spatial part yields the damping force of (\ref{al}) in the non-relativistic
limit, is known to arise from the effect of the charge's `near' (as opposed to radiation) field \cite{teitelboim,rohrlich}.

The appearance of a third-order term in (\ref{al}) and (\ref{ald}) leads to unphysical
behavior, including pre-accelerating and self-accelerating (or `runaway')  solutions. These deficiencies are known to
originate from the assumption that the charge is pointlike. For a non-relativistic charge distribution of
small but finite
size $\ell$,  (\ref{al}) is corrected by an infinite number of higher-derivative terms,
\begin{equation}\label{alextended}
m\left({d^2 \vec{x}\over dt^2}-t_e\left[{d^3 \vec{x}\over dt^3}-\sum_{n=1}^{\infty} b_n {\ell^n\over c^n}{d^{3+n}\vec{x}\over dt^{3+n}}\right]\right)=\vec{F}~
 \end{equation}
(with $b_n$ some numerical coefficients), and is physically sound as long as $\ell>c t_e$ \cite{rohrlich}.
Upon shifting attention to the quantum case, one intuitively expects the pointlike `bare' electron to acquire an effective size of order the Compton wavelength $\lambda_C\equiv \hbar/mc$, due to its surrounding cloud of virtual particles. Given that $\lambda_C\gg ct_e$, there should then be no room for unphysical behavior. Indeed, in \cite{monizsharp} it was shown that nonrelativistic QED leads to
\begin{equation}\label{alextendedquantum}
m\left({d^2 \vec{x}\over dt^2}-t_e\left[{d^3 \vec{x}\over dt^3}-\sum_{n=1}^{\infty} d_n {\lambda_C^n\over c^n}{d^{3+n}\vec{x}\over dt^{3+n}}\right]\right)=\vec{F}~,
 \end{equation}
 which has no runaway solutions, and shows that the charge develops a characteristic size $\ell=\lambda_C$.

Going further to the quantum non-Abelian case is a serious
challenge.
Nevertheless, we will  show here that the AdS/CFT correspondence
\cite{malda}
 allows us to address this question rather easily in certain
strongly-coupled non-Abelian gauge theories \cite{lorentzdirac,damping}.
 We expect the basic story we will uncover to apply generally to all instances of the gauge/string duality (including cases with finite temperature or chemical potentials), but  for simplicity we will concentrate on the specific example of $\cN=4$ SYM.
 Besides the gauge field, this maximally supersymmetric and conformally invariant theory (CFT) contains 6 real scalar fields and 4 Weyl fermions, all in the \emph{adjoint} representation of the gauge group.
 We will be able to derive a Lorentz covariant equation summarizing the dynamics of a quark in this strongly-coupled theory, which will turn out to be a nonlinear generalization of the Lorentz-Dirac equation (\ref{ald}), with a higher-derivative structure somewhat similar to (\ref{alextendedquantum}).

\section{Basic Setup}

{}From this point on we work in natural units, $c=\hbar=1$. It is by now well-known that  $\cN=4$ $SU(N_c)$ SYM with
coupling $g_{YM}$ is, despite appearances, completely equivalent \cite{malda} to Type IIB string theory
on a background that asymptotically approaches the five-dimensional anti-de Sitter (AdS) geometry
\begin{equation}\label{metric}
ds^2=G_{MN}dx^M dx^N={R^2\over z^2}\left(
-dt^2+d\vec{x}^2+{dz^2}\right)~,
\end{equation}
where ${R^4/\ls^4}=g_{YM}^2 N_c\equiv\lambda$ denotes the 't Hooft coupling, and $\ls$ is the string length. The
radial direction $z$ is mapped holographically into a variable length scale in the gauge
theory, in such a way that $z\to 0$ and $z\to\infty$ are respectively the ultraviolet and infrared limits. The directions $x^{\mu}\equiv(t,\vec{x})$ are parallel to the AdS boundary $z=0$ and are
directly identified with the gauge theory directions.
The state of IIB string theory described by the unperturbed metric (\ref{metric}) corresponds to the vacuum of the $\cN=4$ SYM theory, and the closed string sector describing (small or large) fluctuations on top of it fully captures the gluonic ($+$ adjoint scalar and fermionic) physics. The string theory description is under calculational control only for small string coupling and low curvatures, which translates into $N_c\gg 1$,  $\lambda\gg 1$.

{}It is also known that one can add to SYM $N_f$ flavors of matter in the \emph{fundamental}
representation of the $SU(N_c)$ gauge group
by introducing in the string theory setup an open string sector associated with a stack of $N_f$ D7-branes \cite{kk}.
 We will refer to these degrees of freedom
as `quarks,' even though, being $\cN=2$ supersymmetric,
they include
both spin $1/2$ and spin $0$ fields.  The D7-brane embedding is chosen to be translationally invariant along the gauge theory directions $x^{\mu}$, and extend in the radial direction  from the boundary of AdS at $z=0$ up to a location
 \begin{equation}\label{zm}
z_m={\sqrt{\lambda}\over 2\pi m}~
\end{equation}
determined by the mass $m$ of the quarks.
 For $N_f\ll N_c$, the backreaction of the D7-branes on the geometry can be neglected; in the field theory this corresponds to
a `quenched' approximation.

An isolated quark of mass $m$ is dual to
an open string that has one endpoint on the D7-branes at $z=z_m$, and extends radially all the way to the AdS horizon at $z\to\infty$. The string dynamics is governed by the Nambu-Goto action
\begin{equation}\label{nambugoto}
S_{NG}=-{1\over 2\pi\ls^2}\int
d^2\sigma\,\sqrt{-\det{g_{ab}}}\equiv \int
d^2\sigma\,\cL_{{ NG}}~, 
\end{equation}
where $g_{ab}\equiv\p_a X^M\p_b X^N G_{MN}(X)$ ($a,b=0,1$) denotes
the induced metric on the worldsheet.
We can exert an external force $\vec{F}$ on the string endpoint by turning on an electric field $F_{0i}=F_i$ on the D7-branes. This amounts to adding to (\ref{nambugoto}) the
usual minimal coupling, which in terms of the endpoint/quark worldline
$x^{\mu}(\tau)\equiv X^{\mu}(\tau,z_m)$ reads
\begin{equation}\label{externalforce}
S_{{ F}}=\int
d\tau\,A_{\mu}(x(\tau))\,{dx^{\mu}(\tau)\over d\tau}~.
\end{equation}
Variation of  $S_{{ NG}}+S_{{ F}}$ implies the standard Nambu-Goto equation of motion for all interior points of the string, plus the boundary condition
\begin{equation}\label{stringbc}
\Pi^{z}_{\mu}(\tau)|_{z=z_m}=\cF_{\mu}(\tau)\quad\forall~\tau~,
\end{equation}
where $\Pi^{z}_{\mu}\equiv {\p\cL_{{ NG}}}/{\p(\p_z X^{\mu})}$
is the worldsheet (Noether) current associated with spacetime momentum, and
$\cF_{\mu}=-F_{\nu\mu}\p_{\tau}x^{\nu}
=(-\gamma\vec{F}\cdot\vec{v},\gamma\vec{F})$
the Lorentz four-force.

Notice that the string is being described (as is customary) in first-quantized
language, and, as long as it is sufficiently heavy, we are allowed to treat it semiclassically.
In gauge theory language, then,
we are coupling a first-quantized quark to the gluonic ($+$ other SYM) field(s), and then carrying out the full path integral over the strongly-coupled field(s) (the result of which is codified by the AdS spacetime), but treating the path integral over the quark trajectory $x^{\mu}(\tau)$ in a saddle-point approximation.\footnote{For a study of quantum fluctuations about this classical string configuration, see the very recent work \cite{brownian}.}

In more detail, it is really the endpoint of the string that corresponds to the quark, while the body of the string codifies the profile of the (near and radiation) gluonic ($+$ other SYM) field(s) set up by the quark.\footnote{In other words, the string here is the $\cN=4$ SYM analog of the `QCD string', with the surprising twist that it lives not in 4 but in 5 ($+5$) dimensions.} The latter can be mapped out explicitly by computing one-point functions of local operators ($\expec{\tr F^2}$,$\expec{T_{\mu\nu}}$, $\ldots$) in the presence of the quark, which, via the standard GKPW recipe \cite{gkpw}, requires a determination of the near-boundary profile of the closed string fields ($\phi$, $h_{\mu\nu}$, $\ldots$) generated by the macroscopic string.

For the interpretation of our results, it will be crucial to keep in mind that the quark described by this string is not `bare' but `composite' or `dressed'. This can be seen most clearly by working out the expectation value of the gluonic field surrounding a static quark located at the origin \cite{martinfsq},
\begin{equation}\label{Fsquared} {1\over 4 g_{YM}^2}\expec{\tr F^2(x)}
={\sqrt{\lambda}\over 16\pi^2|\vec{x}|^4}\left[1-\frac{1+{5\over 2}\left({2\pi
m|\vec{x}|\over\sqrt{\lambda}}\right)^2}{\left(1+\left({2\pi
m|\vec{x}|\over\sqrt{\lambda}}\right)^2\right)^{5/2}}\right]~. \end{equation} For $m\to\infty$
($z_m\to 0$), this is just the Coulombic field expected (by conformal invariance) for a pointlike
charge. For finite $m$, the profile is still Coulombic far away from the origin, but in fact becomes
non-singular at the location of the quark,
\begin{equation} \label{martinfsqsimplified}
{1\over 4 g_{YM}^2}\expec{\tr
F^2(x)}={\sqrt{\lambda}\over 128\pi^2}\left[15\left({2\pi m\over\sqrt{\lambda}}\right)^4-{35\over
|\vec{x}|^4}\left({2\pi m|\vec{x}|\over\sqrt{\lambda}}\right)^6+\ldots\right]~
\;\mbox{for}~\;|\vec{x}|< {\sqrt{\lambda}\over 2\pi m}~.\nonumber
\end{equation}
 As seen in these equations, the characteristic thickness of this non-Abelian charge distribution
 is precisely the length scale $z_m$ defined in (\ref{zm}). This is then the size of
the gluonic cloud that surrounds the quark, or in other words, the analog of the Compton wavelength for our non-Abelian source.

\section{Generalized Lorentz-Dirac Equation for the Quark}

By Lorentz invariance, given the description of the static quark, we know that a quark moving at constant velocity corresponds to a purely radial string, moving as a rigid vertical rod. When the quark accelerates, the body of the string will trail behind the endpoint, and will therefore exert a force on the latter. Remembering that the body of the string codifies the SYM fields sourced by the quark, we know that in the gauge theory this force is interpreted as the backreaction of the gluonic field on the quark. In other words, in the AdS/CFT context the quark has a `tail', and it is this tail that is responsible for the damping effect we are after.
This mechanism had been previously established in the computations of the drag force exerted on the quark by a thermal plasma, which is described in dual language in terms of a string living on a black hole geometry \cite{hkkky,gubser}. Our analysis here will make it clear that the damping effect is equally present in the gauge theory vacuum \cite{lorentzdirac,damping}.

To flesh out this story, we need to determine the string profile corresponding to a given accelerated quark/endpoint trajectory, by solving the nonlinear equation of motion following from the Nambu-Goto action (\ref{nambugoto}).
Fortunately, for this  task we can make use of the results obtained in a remarkable paper by Mikhailov
\cite{mikhailov}, which we now briefly review (a more detailed explanation can be found in
\cite{dragtime}). This author considered an infinitely massive quark ($z_m=0$), and was able to solve the
equation of motion for the dual string on AdS, for an \emph{arbitrary} timelike trajectory of the
string endpoint. In terms of the coordinates used in (\ref{metric}), his solution is
\begin{equation}\label{mikhsol} X^{\mu}(\tau,z)=z{dx^{\mu}(\tau)\over d\tau}+x^{\mu}(\tau)~,
\end{equation} with $x^{\mu}(\tau)$ the worldline of the string endpoint at the AdS boundary--- or,
equivalently, the worldline of the dual, infinitely massive, quark--- parametrized by its proper time $\tau$. It is easy to check that the lines at constant $\tau$ are null with respect to the induced worldsheet metric, a fact that plays an important role in Mikhailov's construction.

The solution (\ref{mikhsol}) is retarded. To see this,
note that, parametrizing the quark worldline by $x^0(\tau)$
instead of $\tau$, and using $d\tau=\sqrt{1-\vec{v}^{\,2}}dx^0$, where $\vec{v}\equiv d\vec{x}/dx^0$, the $\mu=0$ component of (\ref{mikhsol}) takes the form
 \begin{equation}\label{tret}
t=z{1\over\sqrt{1-\vec{v}^{\,2}}}+\tret~, \end{equation}
where $\tret$ denotes the value of the quark/endpoint coordinate time $x^{0}(\tau)$ at the proper time $\tau$ relevant to the $(t,z)$ segment of the string under consideration, and the endpoint velocity $\vec{v}$ is
meant to be evaluated at $\tret$. In these same terms, the spatial components of (\ref{mikhsol}) can
be formulated as
\begin{equation}\label{xmikh}
\vec{X}(t,z)=z{\vec{v}\over\sqrt{1-\vec{v}^{\,2}}}+\vec{x}(\tret)=(t-\tret)\vec{v}+\vec{x}(\tret)~.
\end{equation}
We see here that the behavior at time $t=X^{0}(\tau,z)$ of the string segment located at radial position $z$ (which, roughly speaking, codifies the gluonic field profile at length scale $z$ and position $\vec{x}=\vec{X}(t,z)$ in the gauge theory) is
completely determined by the behavior of the string endpoint at the \emph{earlier} time $\tret(t,z)$ determined from (\ref{tret}), i.e.,
 by projecting back toward the boundary along the null line at fixed $\tau$. (Note the analogy with the construction of the Lienard-Wiechert fields in classical electromagnetism.)

 An analogous
advanced solution built upon the same endpoint/quark trajectory can be obtained by reversing the
sign of the first term in the right-hand side of (\ref{mikhsol}). In gauge theory language, this
choice of sign corresponds to the choice between a purely outgoing or purely ingoing boundary
condition for the waves in the gluonic field at spatial infinity. Both on the string and the gauge
theory sides, more general configurations should of course exist, but obtaining them explicitly is
difficult due to the highly nonlinear character of the system. Henceforth we will focus solely on
the retarded solutions, which are the ones that capture the physics of present interest, with
influences propagating outward from the quark to infinity.

Using (\ref{tret}) and (\ref{xmikh}), Mikhailov was able to rewrite the total string
energy in the form \begin{equation}\label{emikh} E(t)={\sqrt{\lambda}\over
2\pi}\int^t_{-\infty}d\tret
\frac{\vec{a}^{\,2}-\left[\vec{v}\times\vec{a}\right]^2}{\left(1-\vec{v}^{\,2}\right)^3}
+E_q(\vec{v}(t))~, \end{equation} where of course $\vec{a}\equiv d\vec{v}/dx^0$.  The second term in the above equation arises from a total derivative on the string
worldsheet, and gives the expected Lorentz-covariant expression for the energy intrinsic to the quark
\cite{dragtime}, \begin{equation}\label{edr} E_q(\vec{v})={\sqrt{\lambda}\over
2\pi}\left.\left({1\over\sqrt{1-\vec{v}^{\,2}}}{1\over z}\right)\right|^{z_m=0}_{\infty}=\gamma m~.
\end{equation}
The first term in (\ref{emikh}) must then represent
 the accumulated energy \emph{lost} by the quark over all times prior to $t$. Surprisingly,
the rate of energy loss is seen to have precisely the same form as the standard Lienard formula from classical
electrodynamics. We therefore learn that   in this non-Abelian, strongly-coupled theory,  the energy loss of an infinitely massive (pointlike) quark depends \emph{locally} on the quark worldline.
For the spatial momentum, \cite{mikhailov,dragtime} similarly find
\begin{equation}\label{pmikh} \vec{P}(t)={\sqrt{\lambda}\over 2\pi}\int^t_{-\infty}d\tret
\frac{\vec{a}^{\,2}-\left[\vec{v}\times\vec{a}\right]^2}{\left(1-\vec{v}^{\,2}\right)^3}
\vec{v}+\vec{p}_q(\vec{v}(t))~, \end{equation} with \begin{equation}\label{pdr}
\vec{p}_q={\sqrt{\lambda}\over 2\pi}\left.\left({\vec{v}\over\sqrt{1-\vec{v}^{\,2}}}{1\over
z}\right)\right|^{z_m=0}_{\infty}=\gamma m\vec{v}~. \end{equation} We see then that, in spite of the
non-linear nature of the system, Mikhailov's procedure leads to a clean separation between the tip
and the tail of the string, i.e., between the quark (including its near field) and its gluonic
radiation field.

In fact, using (\ref{mikhsol}), the standard string dynamics reduces to standard particle dynamics at the level of the action: plugging (\ref{mikhsol}) back into (\ref{nambugoto})$+$(\ref{externalforce}),
we can explicitly carry out the integral over $z$ to obtain \cite{damping}
\begin{eqnarray}\label{relativisticparticle} S_{\mbox{\scriptsize NG}}+S_{\mbox{\scriptsize
F}}&=&-{R^2\over 2\pi\ls^2}\int d\tau \int_{z_m\to 0}^{\infty} {dz\over z^2}+\int
d\tau\,A_{\mu}(x(\tau)){dx^{\mu}\over d\tau}(\tau)\\ {}&=& -m\int d\tau~+\int
d\tau\,A_{\mu}(x(\tau)){dx^{\mu}\over d\tau}(\tau)~,\nonumber
\end{eqnarray}
which is evidently the
standard action for a pointlike externally forced relativistic particle (with mass $m\to\infty$).
Notice that the associated equation of motion does \emph{not} include a damping force, which is just
as one would expect for an infinitely massive charge, because the coefficient $t_e\propto 1/m$ of the
damping terms in (\ref{al}) and (\ref{ald}) approaches zero as $m\to\infty$.

Let us now consider the more interesting case of a quark with finite mass, $z_m>0$, where there should be a noticeable damping effect. As we
emphasized at the end of the previous section, in this case our non-Abelian source is no longer pointlike but has size
$z_m$. On the string theory side, the string endpoint is now at $z=z_m$, and we must again require it to follow the given quark trajectory, $x^{\mu}(\tau)$. As before, this condition by itself does not pick out a unique string embedding. Just like we discussed for the infinitely massive case, we additionally require the solution to be retarded, in order to focus on the gluonic field causally set up by the quark. As in \cite{dragtime}, we can inherit this structure by truncating a suitably selected retarded Mikhailov solution.
The embeddings of interest to us can thus be regarded as the $z\ge z_m$ portions of the solutions (\ref{mikhsol}). These are parametrized by data at the AdS boundary $z=0$, which  are now merely auxiliary and  will  henceforth be denoted with tildes, to distinguish them from the actual physical data associated with the endpoint/quark at $z=z_m$.
In this notation, the string embedding reads
\begin{equation}\label{mikhsoltilde}
X^{\mu}(\ttau,z)=z{d\tx^{\mu}(\ttau)\over d\ttau}+\tx^{\mu}(\ttau)~.
\end{equation}
Differentiation with respect to $\ttau$ and evaluation at $z=z_m$ (where we can read off the quark trajectory $x^{\mu}(\ttau)\equiv X^{\mu}(\ttau,z_m)$) leads to
\begin{equation}\label{xdot}
{d x^{\mu}\over d\ttau}=z_m{d^{2}\tx^{\mu}\over d\ttau^{2}}+{d\tx^{\mu}\over d\ttau}~,
\end{equation}
which in turn implies
\begin{equation}\label{tau}
d\tau^2\equiv -dx^{\mu}dx_{\mu}=d\ttau^2\left[1-z^2_m\left({d^2\tx\over d\ttau^2}\right)^2\right]~
\end{equation}
and
\begin{equation}\label{xdotdot}
{d^2 x^{\mu}\over d\ttau^2}=z_m{d^{3}\tx^{\mu}\over d\ttau^{3}}+{d^2\tx^{\mu}\over d\ttau^2}~.
\end{equation}

We are now finally ready to derive the desired equation of motion for the quark. This must simply be dual to the equation of motion satisfied by the string endpoint, which we know to be given by the standard boundary condition (\ref{stringbc}). For the embeddings (\ref{mikhsoltilde}), this condition reads
\begin{equation}\label{stringbcmikh}
\Pi^{z}_{\mu}(\tau)
={\sqrt{\lambda}\over 2\pi}{d\ttau\over d\tau}\left[{1\over z_m}{d^2 \tx^{\mu}\over d\ttau^2}+\left({d^2 \tx^{\mu}\over d\ttau^2}\right)^2 {d\tx^{\mu}\over d\ttau}\right]
=\cF_{\mu}~.
\end{equation}
 Using (\ref{tau}), (\ref{xdotdot}) and carrying out some additional algebra (see \cite{damping} for details), this can be rewritten in the form
\begin{equation}\label{eom}
{d\over d\tau}\left(\frac{m{d x^{\mu}\over d\tau}-{\sqrt{\lambda}\over 2\pi m} \cF^{\mu}}{\sqrt{1-{\lambda\over 4\pi^2 m^4}\cF^2}}\right)=\frac{\cF^{\mu}-{\sqrt{\lambda}\over 2\pi m^2} \cF^2 {d x^{\mu}\over d\tau}}{1-{\lambda\over 4\pi^2 m^4}\cF^2}~,
\end{equation}
which is our main result \cite{lorentzdirac,damping}.

Notice that the characteristic length scale appearing in (\ref{eom}) is precisely $z_m=\sqrt{\lambda}/2\pi m$, which as discussed below (\ref{martinfsqsimplified}), is the quark Compton wavelength.
Let us now examine the behavior of a quark that is sufficiently heavy, or is forced sufficiently softly, that the condition $\sqrt{\lambda |\cF^2|}/2\pi m^2\ll 1$  holds. It is then natural to expand the equation of motion in a power series in this small parameter. To zeroth order in this expansion, we correctly reproduce the pointlike result
$m \p_{\tau}^2 x^{\mu}=\cF^{\mu}$. If we instead keep terms up to first order, we find
  $$
  m {d\over d\tau}\left( {d x^{\mu}\over d\tau}-{\sqrt{\lambda}\over 2\pi m^2}\cF^{\mu}\right)\simeq\cF^{\mu}-{\sqrt{\lambda}\over 2\pi m^2}\cF^2 {d x^{\mu}\over d\tau}~.
  $$
  In the $\cO(\sqrt{\lambda})$ terms it is consistent, to this order, to replace $\cF^{\mu}$ with its zeroth order value, thereby obtaining
  \begin{equation}\label{ourald}
  m \left( {d^2 x^{\mu}\over d\tau^2}-{\sqrt{\lambda}\over 2\pi m}{d^3 x^{\mu}\over d\tau^3}\right)\simeq\cF^{\mu}-{\sqrt{\lambda}\over 2\pi}{d^2 x^{\nu}\over d\tau^2}{d^2 x_{\nu}\over d\tau^2} {d x^{\mu}\over d\tau}~.
  \end{equation}
  Interestingly, this coincides \emph{exactly} with the Lorentz-Dirac equation (\ref{ald}), with the Compton wavelength (\ref{zm}) playing the role of characteristic size $t_e$ for the composite quark. This is indeed the natural quantum scale of the problem. The radiation reaction force in (\ref{ourald}) is correctly given by the
  covariant Lienard formula, as expected from the  result (\ref{emikh}) \cite{mikhailov}, which we see then arising as
  the pointlike limit of the full radiation rate encoded in the right-hand side of (\ref{eom}).
 The
  Schott term in (\ref{ourald}) (associated with the near field of the quark), originated from the terms inside the $\tau$-derivative in the left-hand side of (\ref{eom}), which we understand then to codify a modified dispersion relation for our composite quark.

To second order in $\sqrt{\lambda |\cF^2|}/2\pi m^2$, we similarly obtain
\begin{eqnarray}\label{ourald2}
m{d^2 x^{\mu}\over d\tau^2}
-{\sqrt{\lambda}\over 2\pi}\left(\overbrace{d^3 x^{\mu}\over d\tau^3}
\underbrace{-{d^2 x^{\nu}\over d\tau^2}{d^2 x_{\nu}\over d\tau^2} {d x^{\mu}\over d\tau}}\right)
&{}&{}\nonumber\\
+{\lambda\over 4\pi^2 m}\left(\overbrace{{d^4 x^{\mu}\over d\tau^4}
-(1}\underbrace{+2){d^2 x^{\nu}\over d\tau^2}{d^3 x_{\nu}\over d\tau^3} {d x^{\mu}\over d\tau}}\right)
&{}&{}\\
-{\lambda\over 4\pi^2 m}\left(\overbrace{1\over 2}+\underbrace{1}\right){d^2 x^{\nu}\over d\tau^2}{d^2 x_{\nu}\over d\tau^2} {d^2 x^{\mu}\over d\tau^2}
&\simeq&\cF^{\mu}~. \nonumber
\end{eqnarray}
For compactness, we have grouped together all terms arising at the same order in the expansion, and used underbraces to mark the radiation reaction terms originating from the right-hand side of (\ref{eom}) (which now include corrections beyond the standard Lienard formula), and overbraces to indicate the near-field terms arising from the left-hand side of (\ref{eom}) (which incorporate corrections to the standard Schott term).

We can continue this expansion procedure to arbitrarily high order in $\sqrt{\lambda |\cF^2|}/2\pi m^2$.
Our full equation (\ref{eom}) is thus recognized as a compact (reduced-order) rewriting of an infinite-derivative extension of the Lorentz-Dirac equation that automatically incorporates the size $z_m$ of our
non-pointlike non-Abelian source. It is curious to note that (\ref{eom}), which clearly incorporates the effect of radiation damping on the quark, has been obtained from (\ref{mikhsol}), which does \emph{not} include such damping for the string itself. The latter arises from the backreaction of the closed string fields set up by our macroscopic string, but these are of order $1/N_c^2$, and therefore
subleading at large $N_c$.

The full physical content of (\ref{eom}) can be made transparent by rewriting it in the form
 \begin{equation}\label{eomsplit}
 {d P^{\mu}\over d\tau}\equiv {d p_q^{\mu}\over d\tau}+{d P^{\mu}_{{ rad}}\over d\tau}=\cF^{\mu},
 \end{equation}
 recognizing $P^{\mu}$ as the total string ($=$ quark $+$ radiation) four-momentum,
 \begin{equation}\label{pq}
 p_q^{\mu}=\frac{m{d x^{\mu}\over d\tau}-{\sqrt{\lambda}\over 2\pi m} \cF^{\mu}}{\sqrt{1-{\lambda\over 4\pi^2 m^4}\cF^2}}
 \end{equation}
 as the intrisic momentum of the quark including the contribution of the near-field sourced by it
 (or, in quantum mechanical language,
  of the gluonic cloud surrounding the quark), and
\begin{equation}\label{radiationrate}
{d P^{\mu}_{{\mbox{\scriptsize rad}}}\over d\tau}={\sqrt{\lambda}\, \cF^2 \over 2\pi m^2}\left(\frac{{d x^{\mu}\over d\tau}-{\sqrt{\lambda}\over 2\pi m^2} \cF^{\mu} }{1-{\lambda\over 4\pi^2 m^4}\cF^2}\right)
\end{equation}
as the rate at which momentum is carried away from the quark by chromo-electromagnetic radiation.
Both (\ref{pq}) and (\ref{radiationrate}) diverge when $\cF^2\to \cF^2_{\mbox{\scriptsize crit}}$, where
  \begin{equation}\label{Fcrit}
  \cF^2_{\mbox{\scriptsize crit}}={ 4\pi^2 m^4\over\lambda}
  \end{equation}
  is the critical value at which the force becomes strong enough to nucleate quark-antiquark pairs
  (or, in dual language, to create open strings) \cite{ctqhat}.

Unlike its classical electrodynamic counterpart  (\ref{ald}), our dressed quark equation of motion has no self-accelerating (runaway) solutions: in the (continuous) absence of an external force, (\ref{eom}) uniquely predicts that the 4-acceleration of the quark must vanish. Interestingly, the converse to this last statement is not true: constant 4-velocity does not uniquely imply a vanishing force. E.g., for motion purely along one dimension, (\ref{eom}) reads
   \begin{equation} \label{alinear}
m a=\frac{F(1-v^2)^{3/2}}{\sqrt{1-{\lambda\over 4\pi^2 m^4}F^2}}
+\frac{{\sqrt{\lambda}\over 2\pi m}{dF\over dt}(1-v^2)}{1-{\lambda\over 4\pi^2 m^4}F^2}~. \end{equation}
For $a=0$, we obtain a differential equation with general solution
$$
F(t)=\pm {2\pi m^2\over \sqrt{\lambda}}\,\mathrm{sech}\!\left[{2\pi\over\gamma\sqrt{\lambda}} m(t-t_0)\right]~,
$$
with $t_0$ an integration constant \cite{damping}. This is clearly non-vanishing for all finite values of $t_0$. We conclude then that the energy provided to the
system by this particular $F(t)$ does not translate into an increase of the quark/endpoint velocity, but into a continuous
modification of the string tail, or, in gauge theory language, a change of the gluonic field profile.
This is again a consequence of the extended, and hence deformable, nature of the quark.

\section{Conclusions}

We have shown how a straightforward analysis of the standard dynamics of a string in AdS neatly captures the physics of a `dressed' or `composite' quark that accelerates in the vacuum of strongly-coupled $\cN=4$ SYM, providing a rather beautiful illustration of the power of the AdS/CFT correspondence. Specifically, we have learned that \cite{mikhailov,dragtime,lorentzdirac,damping}:
\begin{itemize}
\item The total 4-momentum of the string (which is conserved at leading order in $1/N_c$) includes not only the {\bf intrinsic} 4-momentum of the dual quark, but also the accumulated 4-momentum {\bf radiated} by the quark.
\item The usual Nambu-Goto equation and forced boundary condition for the string imply a {\bf generalized Lorentz-Dirac equation} for the dressed quark, which is nonlinear and physically sensible.
\item {\bf Radiation damping} is naturally incorporated in this equation of motion for the quark, even though string damping is negligible (at leading order in $1/N_c$).
\item A quark with finite mass is automatically {\bf non-pointlike}, and has a non-trivial and {\bf non-local} dispersion relation and radiation rate.
\item We expect an {\bf analogous story for other strongly-coupled theories} (conformal or not, and with or without temperature or chemical potentials).
\end{itemize}

\section*{Acknowledgements}

We are grateful to the organizers of the Quantum Theory and Symmetries 6 Conference at the University of Kentucky and the XII Mexican Workshop of Particles and Fields in Mazatl\'an, M\'exico, for putting together very useful meetings, and for the opportunity to present this work, which was partially supported by CONACyT grants 50-155I and CB-2008-01-104649, as well as DGAPA-UNAM grant IN116408.


\vspace{1cm}

\begin{thebibliography}{99}

\bibitem{malda}
  J.~M.~Maldacena,
  ``The large $N$ limit of superconformal field theories and
  supergravity,''
  Adv.\ Theor.\ Math.\ Phys.\  {\bf 2}, 231 (1998)
  [Int.\ J.\ Theor.\ Phys.\  {\bf 38}, 1113 (1999)]
  [arXiv:hep-th/9711200].

\bibitem{gkpw}
  S.~S.~Gubser, I.~R.~Klebanov and A.~M.~Polyakov,
  ``Gauge theory correlators from non-critical string theory,''
  Phys.\ Lett.\ B {\bf 428}, 105 (1998)
  [arXiv:hep-th/9802109];\\
  E.~Witten,
  ``Anti-de Sitter space and holography,''
  Adv.\ Theor.\ Math.\ Phys.\  {\bf 2}, 253 (1998)
  [arXiv:hep-th/9802150].

 \bibitem{magoo}
 O.~Aharony, S.~S.~Gubser, J.~M.~Maldacena, H.~Ooguri and Y.~Oz,
   ``Large $N$ field theories, string theory and gravity,''
  Phys.\ Rept.\  {\bf 323}, 183 (2000)
  [arXiv:hep-th/9905111].

 \bibitem{hkkky}
  C.~P.~Herzog, A.~Karch, P.~Kovtun, C.~Kozcaz and L.~G.~Yaffe,
  ``Energy loss of a heavy quark moving through $\cN = 4$ supersymmetric
  Yang-Mills plasma,''
  JHEP {\bf 0607} (2006) 013
  [arXiv:hep-th/0605158].

\bibitem{gubser}
  S.~S.~Gubser,
``Drag force in AdS/CFT,''
  Phys.\ Rev.\  D {\bf 74} (2006) 126005
  [arXiv:hep-th/0605182].

\bibitem{quark}
  J.~Casalderrey-Solana and D.~Teaney,
  ``Heavy quark diffusion in strongly coupled $\cN = 4$ Yang Mills,''
  Phys.\ Rev.\  D {\bf 74} (2006) 085012
  [arXiv:hep-ph/0605199];
  H.~Liu, K.~Rajagopal and U.~A.~Wiedemann,
  ``Calculating the jet quenching parameter from AdS/CFT,''
  Phys.\ Rev.\ Lett.\  {\bf 97} (2006) 182301
  [arXiv:hep-ph/0605178].

\bibitem{meson}
  K.~Peeters, J.~Sonnenschein and M.~Zamaklar,
  ``Holographic melting and related properties of mesons in a quark gluon
  plasma,''
  Phys.\ Rev.\  D {\bf 74}, 106008 (2006)
  [arXiv:hep-th/0606195];
  H.~Liu, K.~Rajagopal and U.~A.~Wiedemann,
  ``An AdS/CFT calculation of screening in a hot wind,''
  Phys.\ Rev.\ Lett.\  {\bf 98}, 182301 (2007)
  [arXiv:hep-ph/0607062];
  M.~Chernicoff, J.~A.~Garc\'\i a and A.~G\"uijosa,
  ``The energy of a moving quark-antiquark pair in an N = 4 SYM plasma,''
  JHEP {\bf 0609}, 068 (2006)
  [arXiv:hep-th/0607089];
  K.~Dusling, J.~Erdmenger, M.~Kaminski, F.~Rust, D.~Teaney and C.~Young,
  ``Quarkonium transport in thermal AdS/CFT,''
  JHEP {\bf 0810}, 098 (2008)
  [arXiv:0808.0957 [hep-th]].

 \bibitem{draggluon}
  M.~Chernicoff and A.~G\"uijosa,
  ``Energy loss of gluons, baryons and $k$-quarks in an $\cN = 4$ SYM plasma,''
  JHEP {\bf 0702} (2007) 084
  [arXiv:hep-th/0611155].

  \bibitem{alr}
  C.~Athanasiou, H.~Liu and K.~Rajagopal,
  ``Velocity Dependence of Baryon Screening in a Hot Strongly Coupled Plasma,''
  JHEP {\bf 0805}, 083 (2008)
  [arXiv:0801.1117 [hep-th]].

\bibitem{ggpr}
  S.~S.~Gubser, D.~R.~Gulotta, S.~S.~Pufu and F.~D.~Rocha,
  ``Gluon energy loss in the gauge-string duality,''
  JHEP {\bf 0810}, 052 (2008)
  [arXiv:0803.1470 [hep-th]].

\bibitem{dragtime}
  M.~Chernicoff and A.~G\"uijosa,
  ``Acceleration, Energy Loss and Screening in Strongly-Coupled Gauge  Theories,''
  JHEP {\bf 0806}, 005 (2008)
  [arXiv:0803.3070 [hep-th]].

\bibitem{vacinplasma}
  Y.~Hatta, E.~Iancu and A.~H.~Mueller,
  ``Jet evolution in the N=4 SYM plasma at strong coupling,''
  arXiv:0803.2481 [hep-th];
  K.~B.~Fadafan, H.~Liu, K.~Rajagopal and U.~A.~Wiedemann,
  ``Stirring Strongly Coupled Plasma,''
  Eur.\ Phys.\ J.\  C {\bf 61}, 553 (2009)
  [arXiv:0809.2869 [hep-ph]].

 \bibitem{abraham}
  M.~Abraham,
  ``Classical theory of radiating electrons,''
  Ann.\ Physik {\bf 10} (1903) 105.

\bibitem{lorentz}
  H.~A.~Lorentz,
  \emph{The Theory of Electrons and Its Applications to the Phenomena of Light and Radiant Heat},
  2nd ed., (Dover, 1952).

   \bibitem{dirac}
  P.~A.~M.~Dirac,
  ``Classical theory of radiating electrons,''
  Proc.\ Roy.\ Soc.\ Lond.\  A {\bf 167} (1938) 148.

  \bibitem{teitelboim}
  C.~Teitelboim,
  ``Splitting of the Maxwell tensor --- radiation reaction without advanced
  fields,''
  Phys.\ Rev.\  D {\bf 1} (1970) 1572
  [Erratum-ibid.\  D {\bf 2} (1970) 1763].

\bibitem{rohrlich}
  F.~Rohrlich,
 \emph{Classical Charged Particles}, 2nd. ed. (Addison Wesley, Redwood City, California, 1990); 
  ``The dynamics of a charged sphere and the electron,''
  Am.\ J.\  Phys. {\bf 65} (1997) 1051.

\bibitem{monizsharp}
  E.~J.~Moniz and D.~H.~Sharp,
  ``Radiation Reaction In Nonrelativistic Quantum Electrodynamics,''
  Phys.\ Rev.\  D {\bf 15} (1977) 2850.

\bibitem{lorentzdirac}
  M.~Chernicoff, J.~A.~Garc\' \i a and A.~G\"uijosa,
  ``Generalized Lorentz-Dirac Equation for a Strongly-Coupled Gauge Theory,''
  Phys.\ Rev.\ Lett.\  {\bf 102} (2009) 241601
  [arXiv:0903.2047 [hep-th]].

\bibitem{damping}
   M.~Chernicoff, J.~A.~Garc\'\i a and A.~G\"uijosa,
  ``A Tail of a Quark in $\cN=4$ SYM,''
  JHEP {\bf 0909} (2009) 080
  [arXiv:0906.1592 [hep-th]].

\bibitem{kk}
  A.~Karch and E.~Katz,
  ``Adding flavor to AdS/CFT,''
  JHEP {\bf 0206} (2002) 043
  [arXiv:hep-th/0205236].

  \bibitem{brownian}
   E.~C\'aceres, M.~Chernicoff, A.~G\"uijosa and J.~F.~Pedraza,
  ``Quantum Fluctuations and the Unruh Effect in Strongly-Coupled Conformal Field Theories,''
  arXiv:1003.5332 [hep-th].

\bibitem{martinfsq}
  J.~L.~Hovdebo, M.~Kruczenski, D.~Mateos, R.~C.~Myers and D.~J.~Winters,
  ``Holographic mesons: Adding flavor to the AdS/CFT duality,''
  Int.\ J.\ Mod.\ Phys.\  A {\bf 20} (2005) 3428.

\bibitem{mikhailov}
  A.~Mikhailov,
  ``Nonlinear waves in AdS/CFT correspondence,''
  arXiv:hep-th/0305196.

\bibitem{ctqhat}
   J.~Casalderrey-Solana and D.~Teaney,
  ``Transverse momentum broadening of a fast quark in a $\cN = 4$ Yang Mills  plasma,''
  JHEP {\bf 0704} (2007) 039
  [arXiv:hep-th/0701123].



\end{thebibliography}
\end{document}